\documentstyle[12pt,a4,epsf]{article}

\newcommand{\dis}{\displaystyle}

\newcommand{\be}{\begin{equation}}
\newcommand{\ee}{\end{equation}}
\newcommand{\nn}{\nonumber}
\newcommand{\bea}{\begin{eqnarray}}
\newcommand{\eea}{\end{eqnarray}}

\newcommand{\gsim}{\ \rlap{\raise 2pt\hbox{$>$}}{\lower 2pt \hbox{$\sim$}}\ }
\newcommand{\lsim}{\ \rlap{\raise 2pt\hbox{$<$}}{\lower 2pt \hbox{$\sim$}}\ }
\newcommand{\ep}{\epsilon_K}
\newcommand{\D}{\Delta}

\newcommand{\ssu}{SU(2)$_L\times$ SU(2)$_R\times$ U(1)$_{B-L}\,$}
\newcommand{\sul}{SU(2)$_L\,$}
\newcommand{\sulu}{SU(2)$_L\times$ U(1)$_Y\,$}
\newcommand{\sur}{SU(2)$_R\,$}

\newcommand{\matr}{\left( \begin{array}}
\newcommand{\ematr}{\end{array} \right)}

\newcommand{\np}[1]{Nucl. Phys. #1}
\newcommand{\pl}[1]{Phys. Lett. #1}
\newcommand{\pr}[1]{Phys. Rev. #1}

\newcommand{\zp}[1]{Z. Phys. #1}

\makeatletter
\if@twoside
\else \oddsidemargin 00pt \evensidemargin 00pt
\fi
\let\@eqnsel = \hfil

\expandafter\ifx\csname mathrm\endcsname\relax\def\mathrm#1{{\rm #1}}\fi
\@ifundefined{mathrm}{\def\mathrm#1{{\rm #1}}}{\relax}

\makeatother

\begin{document}

\begin{titlepage}
\begin{flushright}
FTUV/96-51\\
IFIC/96-59\\
hep-ph/9611347
\end{flushright}

\vskip 1.5cm

\begin{center}
{\Large \bf      
CONSTRAINTS ON THE $W_R$ MASS AND CP-VIOLATION
IN LEFT-RIGHT MODELS \par} \vskip 2.em
{\large		
{\sc Gabriela Barenboim$^{1,2}$, Jos\'e Bernab\'eu$^1$, 
Joaquim Prades$^{1,2},$ and Martti Raidal$^{1,2}$
}  \\[1ex] 
{\it $^1$ Departament de F\'\i sica Te\`orica, Universitat 
de Val\`encia}\\ 
{\it C/ del Dr. Moliner 50, 
E-46100 Burjassot (Val\`encia), Spain.} \\
{\it $^2$ IFIC, Universitat de Val\`encia - CSIC} \\
{\it C/ del Dr. Moliner 50, 
E-46100 Burjassot (Val\`encia), Spain.} \\[1ex]
\vskip 0.5em
\par} 
{\bf Abstract} \end{center} 
We update the constraints on the right-handed $W_R$ gauge
 boson mass, mixing angle $\zeta$ 
with the left-handed $W_L$ gauge boson, and other parameters
in general
left-right symmetric models with different mechanisms of CP 
violation. Constraints mostly independent of any assumption on the quark 
sector are obtained from a re-analysis of muon decay data.
The best $\chi^2$ fit of the data gives $g_R/g_L=0.94\pm 0.09$
for the ratio of right to left gauge couplings, 
with $M_{W_R} \geq 485$ GeV and $\mid \zeta \mid \leq 0.0327$.
Fixing $g_L=g_R$ (in particular for manifestly left-right
 symmetric models), we obtain $M_{W_R}\gsim 549$ GeV 
and $ \mid \zeta\mid\lsim 0.0333$. 
Estimates of the left-right  hadronic matrix elements in the neutral kaon 
system and their uncertainties 
are revised using large $N_c$ and Chiral Perturbation Theory
arguments. With explicitly given  assumptions 
on the long-distance $(\Delta S=1)^2$ contributions
to the $K_L-K_S$ mass difference, 
lower  bounds on $M_{W_R}$ are obtained. With the same assumptions, one 
also gets strong upper bounds from the CP-violating parameter
$\epsilon_K$, for most of the parameter 
space of left-right models where the right-handed third family does not  
contribute in CP-violating quantities. 
For manifestly left-right symmetric models 
the lower bound obtained is $M_{W_R}\gsim(1.6^{+1.2}_{-0.7})$ TeV.  
\par
\vfill
November 1996 
\end{titlepage}

\section{Introduction} 
\label{first}

 While the Standard Model (SM) has been
 successful in its predictions over the past decades it is still
 not fully satisfactory in many ways. The origin of
the maximal parity violation in the weak interactions, 
 the origin of CP-violation and 
the smallness of the ratio of neutrino masses to the top-quark mass
are among the 
 open questions which motivate searches for new physics 
beyond the electroweak scale.
All these puzzles 
find natural answers in extensions of the SM based on
the gauge   group SU(2)$_L$ $\times$ SU(2)$_R$ $\times$
 U(1)$_{B-L} $ \cite{lr,neutrino}.
The most definite benchmark of this class of models would be the discovery
of the  right-handed  currents predicted by them.

Manifestations of  right-handed charged currents
have been looked for, both in high- and low-energy experiments.
Direct searches at Tevatron set 652 GeV \cite{tev} as the lower bound 
of the right-handed gauge boson mass. If the right-handed neutrino 
is assumed to be much lighter than the $W_R$ boson 
 then this bound is increased to  720 GeV \cite{tev2}.
The contribution of virtual $W_R$ excitations 
 in low-energy processes can be used to constrain its
mass, coupling, and other parameters too.
In fact, so far, the most sensitive probe to the additional right-handed 
interaction is provided by the system of neutral kaons \cite{beall}.
The right-handed charged current can give substantial contributions
to  strangeness changing in two units effective Lagrangian which governs the 
$K^0-\overline K^0$ mixing. Thus, well measured
observables of the neutral kaon system  as
the $K_L-K_S$ mass difference $\Delta m_K \equiv m_{K_L}-m_{K_S}$ 
and the CP violating parameter $\epsilon_K$, are most suited for indirect 
limits on the right-handed gauge boson couplings.

Constraints on the 
$W_R$ mass and its mixing angle with the left-handed 
$W_L$ gauge boson 
have been studied, in  general left-right (LR) models, 
 extensively in \cite{uma}.
However, we think that some additional analyses 
might be of interest.
Namely, the CP breaking mechanism effects have not been
fully  exploited in these analyses. 
Because the number of CP violating phases,
$[(N-1)(N-2) + N(N+1)]/2$ in a $N$-family
LR model\footnote{As it is well known, CP-violation can occur even 
in the two family case.}, the effects of CP-violation can 
be expected to be more important than in the SM. In particular,
CP-violating phases can also modify CP-conserving observables. 
For example,
the contributions to $\Delta m_K$ in LR models
are proportional to cosines of  differences of  CP-violating phases  
 which, in the general case, can be arbitrary and
therefore reduce the limits considerably.

 Moreover,  the experimental value of $\ep$,
which in some particular cases  ( e.g. manifestly LR models with 
spontaneous  breaking of CP) has shown to be 
very constraining \cite{meie},   
has not been used to constrain  other left-right models.
We will show that indeed $\ep$ sets very constraining
bounds in a large class of left-right models.
Another point we would like to re-analyse is
the reliability of the LR hadronic matrix elements estimates
in the literature. In particular, we would  like to make a realistic
estimate of its uncertainty at present.
This is necessary in order to have a meaningful
comparison between  the constraints obtained from the neutral kaon  system
and other results, 
e.g. muon decay data or collider experiments.

In this paper we update bounds on  $M_{W_R}$ and 
its mixing angle, $\zeta$,  in general
LR symmetric models with different discrete symmetries
on the Lagrangian. Bounds independent 
on the  quark sector and hadronic physics uncertainties
are set  using updated electroweak data on the
muon decay.  Here we assume that the right-handed
neutrino is light enough to be produced in this decay. 
In the kaon system, we estimate the LR hadronic matrix elements 
and their present uncertainty
using large $N_c$ and chiral symmetry arguments.
We derive bounds on $M_{W_R}$ 
from measurements of both 
$\Delta m_K$ and $\epsilon_K$ in models
with different CP breaking mechanisms and compare
them with results from other sources.

As it has been pointed out by several authors \cite{uma,rizzo,other}, 
bounds on particular $W_R$ parameters like its mass,
depend strongly on theoretical assumptions about
the size of the right-handed gauge coupling $g_R$ and/or 
 the right-handed Cabibbo-Kobayashi-Maskawa (CKM) matrix elements.
Combining the bounds obtained from the two neutral
kaon observables we can eliminate one of the unknowns.

In the Standard Model,
 all three families need to be involved in CP-violation
observables. In addition, there is no CP-violation 
if the up-type quarks or the down-type quarks
are degenerated in mass \cite{jarlskog},
so that one can neglect the light-quark contributions
owing to the large top-quark mass in observables
like $\epsilon_K$.  Observables like
$\Delta m_K$, only sensitive to
the real part of the Lagrangian can, however, get contributions
from each family separately. 
The $K_L-K_S$ mass difference has
been estimated in  the SM in \cite{klks} by matching short- and 
long-distance contributions in a $1/N_c$ expansion
($N_c$ is the number of QCD colors).
 The result indicates that the long-distance
contributions in $\Delta m_K$
in the SM are of the order of 50$\%$.
One thus expects this same large long-distance
QCD contribution to appear in CP-violating observables
when only the
two lightest families are involved, as happens in 
general in LR models. More comments on this issue
are in Section \ref{fourth}.

The outline on the paper is the following.
In Section \ref{second} we briefly 
present the general structure of the left-right
 models we are interested in. In Section \ref{third}  
 we carry out the analysis of constraints from the muon decay data and 
in Section \ref{fourth} we discuss  the effective Lagrangian with 
$\Delta S=2$ in  LR models obtaining constraints on the $W_R$
mass  from $\D m_K$ and $\ep$ measurements. 
In these two sections, we put some emphasis in giving 
explicitly which have been, in each
case, the assumptions and/or the range of applicability of 
our  results. Our conclusions are given in Section \ref{conclusions}.

\section{Left-Right Symmetric Models and CP-Violation} 
\label{second}

Here we  present the basic structure of the \ssu
left-right symmetric model we use in our analyses.
 Since we are interested in the charged current processes
and  CP-violation effects, we concentrate on the gauge and Higgs 
sectors of the model. In left-right models, each familiy 
of fermions 
\bea
\Psi(x) \equiv  {\matr{cc} \nu & u \\ l & d \ematr}^a \, 
\hspace*{1cm} .
\eea
is assigned to doublets of the 
gauge groups SU(2)$_L$ and SU(2)$_R$ according to
the chirality. The Latin index $a=1, \cdots, N$ is for the family. 
Here and in the rest of the paper,
$\Psi_{L(R)}\equiv\left(\frac{\displaystyle 1-(+)\gamma_5}
{\displaystyle 2} \right) \Psi$.
The field $\Psi_L(x)$ transforms under \sul $\times$ \sur 
gauge rotations as (2,1) and the field $\Psi_R(x)$
as (1,2), where the representations are identified by 
their
dimension. Quarks have $B-L=1/3$ and leptons $B-L=-1$.
Their interactions with the corresponding charged gauge bosons 
are determined by the charged current Lagrangian
\bea
{\cal L}_{cc}&=&\sum^{N}_{a=1}\left[\frac{g_L}{\sqrt{2}}
W_L^{\mu}\left(\bar{l}^a_{L}\gamma_{\mu}\nu^a_{L} + 
 \bar{d}^a_{L}\gamma_{\mu}u^a_{L}
 \right) \right.  \nn \\ 
&+& \left. 
\frac{g_R}{\sqrt{2}} W_R^{\mu}
\left( \bar{l}^a_{R}\gamma_{\mu}\nu^a_{R}  + 
 \bar{d}^a_{R}\gamma_{\mu}u^a_{R}  \right)\right] + h.c.
\eea 
The minimal set of fundamental
scalars consists of a bi-doublet $\phi(x)$, and
a left-handed and a right-handed multiplets of Higgs boson fields. 
If one wants to realise the see-saw mechanism \cite{neutrino,seesaw} 
the latter ones 
should be chosen to be triplets $\D_L$ and $\D_R$. 
In this case 
the model contains the following set of Higgs fields,  
\be
\begin{array}{c} {\phi \equiv
\matr{cc}\phi_1^0&\phi_1^+\\\phi_2^-&\phi_2^0
\ematr}, \;\;\;\;
 {\Delta_{L(R)} \equiv 
\matr{cc}\Delta^+&\sqrt{2}\Delta^{++}\\ 
\sqrt{2}\Delta^0&-\Delta^+ \ematr}_{L(R)}.
\end{array}
\ee 
The field matrix  $\phi(x)$ transforms under \sul $\times$ \sur
gauge rotations as (2,2), the field matrix $\D_L$ as (3,1)
and the field matrix $\D_R$ as (1,3).  The  $B-L$ charge
is 0 for the bi-doublet and  2 for the triplets. 
With this field content,  
the most general form of the scalar potential $V(\phi,\D)$
can be found in the literature \cite{desp}.
Sometimes the full Lagrangian of the theory 
is required to be invariant under the transformations
\be
\Psi_L \longleftrightarrow \Psi_R \;,\;\; 
\Delta_L \longleftrightarrow \Delta_R  \;,\;\;
\phi \longleftrightarrow \phi^\dagger.
\label{trans}
\ee
These are the so-called manifestly LR symmetric models,
in this case one also has $g_L=g_R.$

Spontaneous symmetry breaking is parametrized by the following
vacuum expectation values (VEVs) of the scalar  fields, 
\be
\begin{array}{c}
{\langle\phi\rangle=\frac{\displaystyle 1}{\displaystyle 
\sqrt{2}}\matr{cc}\kappa_1&0\\0&\kappa_2\ematr,}
\;\;\;\;
{\langle\Delta_{L,R}\rangle
=\frac{\displaystyle 1}{\displaystyle \sqrt{2}}\matr{cc}0&0\\v_{L,R}&0
\ematr.}
\end{array}
\ee  
The VEVs of the bi-doublet $\phi(x)$
parametrize the spontaneous symmetry breaking of the 
SM gauge group \sulu, they   generate Dirac mass
terms to fermions through the Yukawa Lagrangian  
\bea
{\cal L}_Y &=& \bar{\Psi}_L f  \phi  \Psi_R \; + 
\bar{\Psi}_L h \tilde{\phi}  \Psi_R \; +  \; h.c. ,
\label{yuk}
\eea
where $f$ and $h$ are matrices in family space collecting the Yukawa  
couplings for both quarks and leptons,
and $\tilde{\phi}\equiv \tau_2 \, \phi^* \tau_2$, with $\tau_2$
the second Pauli matrix. Summation over families
is understood.  Diagonalization of the 
up- and down-quark mass matrices in (\ref{yuk}) provides  us with the 
CKM matrices $K_L$ and $K_R$ which, in general, are different.
Analogously, we get the CKM-like matrices for the leptons, 
$U_L$ and $U_R$.
The VEVs $\kappa_1$ and $\kappa_2$
 give also mass to the left-handed  $W_L$ gauge boson.
The left-handed triplet VEV $v_L$ does not play any dynamical 
r\^ole in the symmetry breaking  and
is forced to be small because of its contribution to the $\rho$ parameter 
\cite{desp,meie2}.
The right-handed triplet $\Delta_R$ breaks the SU(2)$_R\times$
U(1)$_{B-L}$ gauge group to
U(1)$_Y$ and its VEV, $v_R$, gives mass to $W_R.$ 
In general, the charged gauge bosons mass eigenstates
are mixings of the flavour eigenstates. The mixing 
angle is
\be
\zeta\equiv\left(\frac{2r}{1+r^2}\right) \, \beta ,
\ee
where $r\equiv\kappa_1/\kappa_2$ and $\beta \equiv 
M_{W_L}^2 / M_{W_R}^2$. 
Just for notation purposes, we will continue to use $W_L$ and $W_R$
for the charged gauge bosons eigenstates, assigning $W_L$ 
to the eigenstate that reduces to the left-handed gauge boson
when $\zeta=0$ and analogously for $W_R$.

There are two natural ways to obtain  breaking of the CP
symmetry  in left-right  models.
Firstly, CP is violated if the Yukawa coupling matrices  
$f$ and $h$ are  complex. This is called
hard CP-violation and is the analogue of the CKM
CP-violation in the SM. 
Secondly, in LR models one can naturally extend the 
idea of  spontaneous breaking of parity 
\cite{app} to the spontaneous violation of CP 
\cite{eg,frere}. This is parametrized by the
 VEVs $v_L$, $v_R$ and
$\kappa_1$, $\kappa_2$, which can, in principle,
 be complex and break CP.
In general left-right models, CP-violation can occur 
either due to just one of the mechanisms or to 
the combination of both.

\section{Updated Constraints from Muon Decay Data} 
\label{third}

Pure leptonic processes are free of both  the assumptions on 
the unknown quark mixings and the 
uncertainty induced by our present limited knowledge on the 
low energy QCD dynamics. They are thus, in principle,
 better suited for obtaining more model independent
constraints. With such aim, we use the updated electroweak data
\cite{pdb}  to re-analyse the muon decay data in the case
of interest here.

As in any model in which light fermions have heavy boson 
mediated interactions, the low energy effective action
of LR symmetric models contains the usual Standard Model
 bilinear terms plus four-fermion interactions. The latter
are the result of integrating out the heavier LR degrees of
freedom. Hence, precise low-energy tests of the light 
fermions constitute a window into the high energy
behaviour of the model underlying the SM (in this
case LR symmetric models).
As was stated before, this procedure is cleaner in the leptonic
sector where hadronization does not obscure it.

The effective Lagrangian which describes the 
contact four-fermion lepton-lepton interaction  
in LR models is \cite{uma}
\bea
{\cal L}_{\mbox{eff}} = g_{LL} \, J_L^+ J_L^- + 
g_{LR} \, J_R^+ J_L^- + g_{RL} 
\, J_L^+ J_R^- + g_{RR} \, J_R^+ J_R^-\;,
\label{ffi}
\eea
with
\bea
J_{L(R)}^{-\mu} \equiv \overline{N}_{L(R)}
 \gamma^\mu \, U_{L(R)} \, E_{L(R)}  \; ,
\eea
where the three neutrino and charged 
lepton families are collected in the 
$N$ and $E$ vectors, respectively and 
\bea
J^{+\mu}_{L(R)} = 
\left( J^{-\mu}_{L(R)}\right)^\dagger \, .
\eea
We assume that right-handed neutrinos are light enough
to be produced in muon decays.
With the present bounds on the left-handed 
neutrino masses \cite{pdb}
the left-handed $U_L$ mixing matrix can be chosen
to be diagonal. 
The couplings in (\ref{ffi}) fulfil $g_{ij}=g^*_{ji}$
with
\bea
g_{LL}& = &\frac{g_L^2}{2 M_{W_L}^2} 
\left( \cos^2\zeta + \beta \sin^2\zeta 
\right), \nonumber \\
g_{LR}& = &  \frac{g_L ^2}{2 M_{W_L}^2} \alpha \left( 1 -
\beta \right) \sin\zeta \cos\zeta e^{i \varphi}, \nonumber \\
g_{RR}& = &\frac{g_L^2}{2 M_{W_L}^2} \alpha^2 
\left( \sin^2\zeta + \beta \cos^2\zeta 
\right) ,
\eea
the angle $\varphi$ is the relative phase of the bi-doublet VEVs
and  $\alpha \equiv g_R/g_L.$
>From $W_L$ gauge boson and $\beta$-decays, 
we know that the leptonic $W_L$
vertices give the dominant contribution
to the muon and tau meson leptonic decays. The SM predicts that this
is indeed the unique tree-level contribution, then in the SM, 
to a very good approximation
\bea
g_{LL} \simeq \frac{g_L^2}{2 M_{W_L}^2} ,
\eea
and $g_{LR}=g_{RR}\simeq 0$. In general LR models with  
the low-energy effective Lagrangian structure in  (\ref{ffi}),
the measured muon decay width puts the following constraint
\bea
8 G_F^\mu = 
{g_{LL}}^2 + 2 \left| g_{LR} \right|^2 + {g_{RR}}^2.
\eea
>From here we can find the relation between the Fermi constant
$G_F^\mu$,  $\alpha$, $\beta$ and the mixing $\zeta$ 
\bea
\frac{G_F^\mu}
{\sqrt{2}} & = & \frac{e^2}
{8 \sin^2\theta_W (1-\Delta r)M_{W_L}^2} A,
\eea
with $|e|$ the electron electric charge and
\bea
\label{adef}
A &=& \left( 1 + \alpha^4 \beta^2 \right)
\cos^4\zeta + \left( \alpha^4 + \beta^2 \right) \sin^4\zeta   
\nonumber \\
&+&2 \left[ \beta \left( 1-\alpha^2 \right)^2 + \alpha^2 \left(
1 + \beta^2 \right) \right] 
\sin^2\zeta \cos^2\zeta ,
\eea
where we applied the one-loop Standard Model radiative 
correction $\Delta r$ to the
fine structure constant. The radiative correction
$\Delta r$ is evaluated numerically for 
$m_t = (175 \pm 6)$ GeV  pole top-quark mass value  
\cite{topmass}  and
$m_H= 300$ GeV yielding $\Delta r = 0.053 \pm 0.003$.
Since we assume that the
SM provides the dominant contribution, any additional 
non-SM higher order correction is 
a sub-leading effect. Therefore only SM radiative corrections
are included.

To determine  constraints on the parameters of 
our LR symmetric model, we have 
expressed the muon decay parameters in terms of
$\alpha$, $\zeta$ and $\beta$ as follows
\bea
\rho &=& \frac{3}{4 A} \left( \left( 1 + \alpha^4 \beta^2 \right)
\cos^4\zeta + \left( \alpha^4 + \beta^2 \right) \sin^4\zeta   
\right. \nonumber \\
&+& \left. 2 \beta \left( 1 + \alpha^2 \right) \sin^2\zeta \cos^2\zeta \right)
,\nonumber \\
\xi &=& -\frac{1}{A} \left( \left( \alpha^4 \beta^2 - 1 \right)
\cos^4\zeta + \left( \alpha^4 - \beta^2 \right) \sin^4\zeta  
\right.  \nonumber \\ &+& \left. 
 2 \beta \left( \alpha^2 - 1 \right) \sin^2\zeta \cos^2\zeta \right),
\nonumber \\
\xi^\prime &=& \xi, \nonumber \\
\xi^{\prime \prime} &=&  \frac{1}{A}
\left( \left( 1 + \alpha^4 \beta^2 \right)
\cos^4\zeta + \left( \alpha^4 + \beta^2 \right) \sin^4\zeta   
\right. \nonumber \\ &+& \left.
2 \left( \beta + 3 \alpha^2 + 3 \alpha^2 \beta^2 - 5 \alpha^2 \beta 
\right) \sin^2\zeta \cos^2\zeta \right),
\nonumber \\
\delta &=& \frac{3}{4} ,\nonumber \\
\overline{\eta} &=& \frac{2}{A} \alpha^2 \left( 1 -\beta \right)^2
\sin^2\zeta \cos^2\zeta,
\eea
where the overall normalization $A$ is given in (\ref{adef}).

We have performed a best $\chi^2$ fit of 
these parameters with the experimental data given by the Particle
Data Group \cite{pdb}. This gives lower limits on $M_{W_R}$ and
upper limits on the mixing angle $\zeta$ for different values of $\alpha$
 as shown in  Table \ref{table1}.
\begin{table}
\begin{center}
\begin{tabular}{||c|c|c||} \hline
    $\alpha=g_R/g_L$      & $M_{W_R}$ (GeV) & $|\zeta|$ \\ \hline
0.50 & $\geq$ 286 & $\leq$ 0.0324 \\ \hline
0.75 & $\geq$ 379 & $\leq$ 0.0321 \\ \hline
1.00 & $\geq$ 549 & $\leq$ 0.0333 \\ \hline
1.50 & $\geq$ 825 & $\leq$ 0.0330 \\ \hline
2.00 & $\geq$ 1015 & $\leq$ 0.0327 \\ \hline
\end{tabular}
\end{center}
\caption{Constraints on the right-handed  $W_R$ gauge boson mass
and mixing angle $\zeta$
for different values of $\alpha=g_R/g_L.$
The second and third columns 
are the corresponding lower limits on $M_{W_R}$
and upper limits on $|\zeta|$, respectively.}
\label{table1}
\end{table} 
These bounds are stronger than the ones obtained in 
previous analyses due to the improvement in 
experimental data precision.
Letting $\alpha$ to vary freely, 
the best chi squared is obtained for  
$\alpha=0.94\pm 0.09$, with 
\bea
M_{W_R} \geq 485 \; \mbox{GeV}\,, \;\;\;\;\;\;\;
\mid \zeta \mid \leq 0.0327 \,.
\eea

\section{Bounds on $M_{W_R}$ in CP Violating Left-Right Models} 
\label{fourth}

As in the SM, in left-right models there are two type of contributions
to strangeness-changing in two units processes, namely 
$\Delta S=2$ transitions induced by box-diagrams \cite{gale,giwi} or 
short-distance contributions, and $(\Delta S=1)^2$ 
transitions or long-distance contributions \cite{longmk}.
As noticed first in \cite{longmk} for the real part and
in \cite{longeps} for the imaginary one, the long-distance
$(\Delta S=1)^2$ contributions can be large.
Its possible importance in LR models was 
already pointed out in \cite{eg,frere}.
Their calculation involves
a good mastering of the QCD long-distance part and in particular
of the so-called $\Delta I=1/2$ rule for $K\to \pi \pi$ decays.
As mentioned in the Introduction, its relative 
importance in CP-violating observables in general LR
models can be larger than in the SM and of the same
order as for CP-conserving quantities like $\Delta m_K$.
This is because, in general LR models, the presence of 
all the SM families is not anymore required in CP-violating
observables and the top-quark contribution
will not dominate in general. One thus expect long- and 
short-distance processes to contribute with the same weight
in CP-conserving and CP-violating observables.
To have a hint, we can compare with what happens in the SM. 
The short- and long-distance contributions
to the CP-conserving observable $\Delta m_K$,  
were computed in \cite{klks} within the SM. 
The result was that 
both contributions are of the same size when the 
scale separating both regions is around 1 GeV.

Therefore, there could be large cancellations
in general LR models between long- and  
short-distance contributions. The precise
analysis requires a careful study of the 
relative phases  and of the LR hadronic matrix elements
of $\Delta S=1$ transitions. This is
 outside the scope of this paper and will be
presented elsewhere.

However, with some more or less strong
assumptions on the long-distance contributions, 
which will be given explicitly in each case, 
one can still
obtain relevant bounds from the short-distance $\Delta S=2$
contributions. 

There already exists an extensive 
literature on the box-diagram contributions
in LR symmetric models \cite{beall,eg,frere,chang}
(for the dominant contributions see Figure 1).
\begin{figure*}[hbtp]
\begin{center}
\mbox{\epsfxsize=16cm\epsfysize=6cm\epsffile{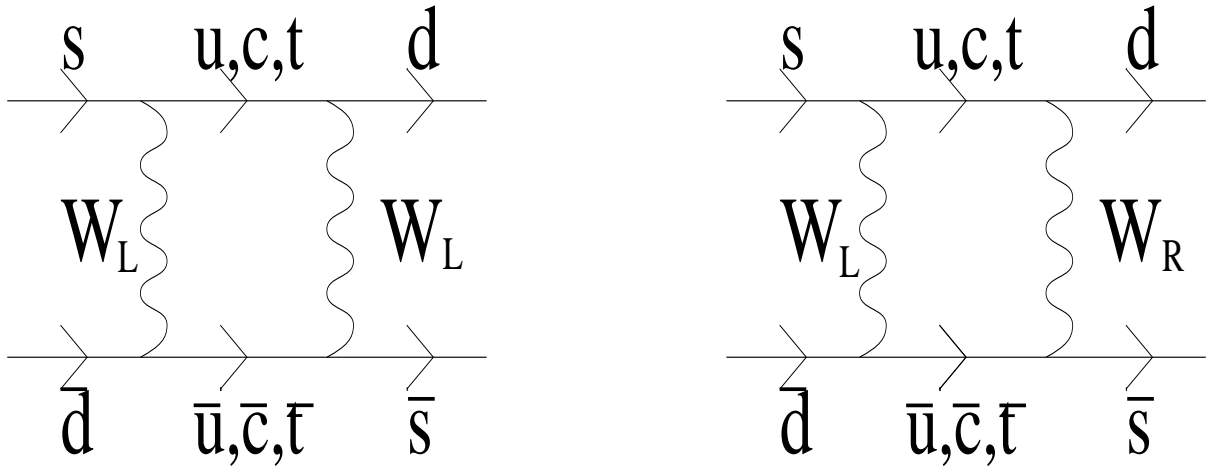}}
\caption{Diagrams showing the dominant box-diagrams
 contributions to the $\Delta S=2$
effective Lagrangian in left-right models.}
\end{center}
\end{figure*}
Namely, the exchange of two left-handed gauge bosons (LL diagram) 
and the exchange of one left and one right-handed
gauge bosons (LR diagrams).  Contributions
coming from right-right gauge boson  
or physical Higgs boson exchanges are suppressed
by boson masses as well as small Yukawa
couplings in the latter case.  
  
For the CKM matrices $K_L$ and $K_R$,
we use the following parametrizations.
In the left-handed sector, we use
a typical SM parametrization of the CKM matrices, 
i.e. three angles and one phase, $\delta$. This can be done
due to our convention in which all 
additional phases are shifted to the right-handed sector. 
For the right handed CKM matrices, the most general matrix has
six phases and three angles. As mentioned in the Introduction,
in general, the right-handed third family  
contribution is not needed for CP-conserving or 
CP-violating observables: this leaves us with three
observable phases,  therefore we can
take the following Wolfenstein-like parametrization
for the lightest two right-handed families of quarks 
sub-matrix, 
\bea
e^{i \gamma} \pmatrix { e^{-i \delta_2} 
\left(1-\frac{\dis \lambda_R^2}{\dis 2}\right) 
 &  e^{-i \delta_1}  \lambda_R \cr
- e^{i \delta_1} \lambda_R & 
e^{i\delta_2}  \left(1-\frac{\dis \lambda_R^2}{\dis 2}\right)  \cr} ,
\label{km}
\eea
which violates unitarity by terms proportional to $\lambda_R^4$.
This is naturally small in left-right models
if the same hierarchy in the angles
as in the left-handed sector holds. The parameter $\lambda_R$ is  the 
right sector analogous to the Cabibbo angle in the Wolfenstein 
parametrization of the  left-handed CKM matrix  $\lambda 
\simeq |V_{us}| \simeq 0.22$.

As pointed out in \cite{frere}, the charm-charm contribution 
dominates over the top-top and top-charm contributions
in the right-handed sector,
unless a fine-tuning of the parameters is made. Assuming the same
hierarchy in the right-handed CKM angles as the one in
the left-handed ones (notice that they are equal in 
some LR models) and the same order of magnitude 
for the contributing right-handed phases in each type
of box-diagram, this dominance   
can be  quantified in  terms of CKM matrix 
elements and quark  masses by the factor  
$\lambda^8 (m_t^2/m_c^2) \mid \ln(m_t/M_{W_L})/ 
\ln(m_c/M_{W_L})\mid \simeq 0.02$ for the top-top
over the charm-charm contributions and by 
$\lambda^4 m_c/m_t \simeq 10^{-5}$ for the top-charm
over the charm-charm contributions.
 Diagrams with unphysical
scalars are suppressed by a $m_c^2/M_{W_L}^2$ factor. 
This dominance is also true for the contributions
of the left-handed sector to the real
part of the effective $\Delta S=2$ Lagrangian.
However the contribution of the left-handed sector
is dominated by the top-quark instead, 
therefore we keep the full
contribution from the left-handed sector 
to the $\Delta S=2$ effective Lagrangian
(see for instance \cite{BBL95}). The 
effective $\Delta S=2$ Lagrangian in LR models
is thus well approximated
 by 
\bea
{\cal L}_{eff}^{\Delta S=2} &\simeq&
-\frac{G_F^2}{4\pi^2}\, M_{W_L}^2\,   
\nonumber \\ &\hspace*{-1cm}\times\hspace*{1cm}&\hspace*{-1cm}
\left[ \left\{\left(\lambda_c^*\right)_{LL}^{2}  \eta_1(\mu)
S(x_c) \,  + \, \left(\lambda_t^*\right)_{LL}^{2}  \eta_2(\mu)
S(x_t) \,+\,  2 \left(\lambda_c^*\right)_{LL} \,  
\left(\lambda_t^*\right)_{LL}  \eta_3(\mu)\, 
 S(x_c,x_t) \right\} \, {\cal O}_{LL}(x)\, 
\right. \nonumber \\ &\hspace*{-1cm}+\hspace*{1cm}&\hspace*{-1cm}
\left. \, 2 x_c \ln(x_c)  
\frac{M_{W_L}^2}{M_{W_R}^2}\, \frac{g_{R}^2}{g_{L}^2}\, 
\left(\lambda_c^*\right)_{LR} \, \left(\lambda_c^*\right)_{RL} 
\, \left( \eta_4(\mu) \, {\cal O}_{S}(x)\, + \, 
\eta_5(\mu) \, \left[ {\cal O}_{LR}(x)
+\frac{2}{N_c} \, {\cal O}_{S}(x) \right]
\right) \right] \nonumber \\
\label{eff}
\eea
where 
\bea
S(x)=x\left[ \frac{1}{4}+\frac{9}{4}\frac{1}{(1-x)}-
\frac{3}{2}\frac{1}{(1-x)^2} \right] -
\frac{3}{2} \left[ \frac{x}{1-x} \right]^3 \ln x \, ;
\nonumber \\
\mbox{\rm and} \hspace*{1cm} 
S(x,y)=x\left[\ln\left(\frac{y}{x}\right)
- \frac{3y}{4(1-y)}\, \left(1+ \frac{y \ln y}{1-y}\right)
\right]
\eea
with $x_i=m_i^2/M_{W_L}^2$ for $m_i$ the $i$-flavour 
$\overline {MS}$-pole quark-mass. In (\ref{eff}) 
$\left(\lambda_i\right)_{AB}\equiv
\left(K_{A}\right)_{id} \left(K_{B}^*\right)_{is}$ and
\bea
{\cal O}_{LL}(x) &\equiv& \left(\bar{d}_L \, \gamma^\mu \,  
s_L \right) (x)\left(\bar{d}_L \, \gamma_\mu  \, s_L\right) (x) \;\;,\;\;
\nonumber \\
{\cal O}_{LR}(x)&\equiv& \left(\bar{d}_L \, \gamma^\mu \, s_L\right)(x)
 \left(\bar{d}_R \, \gamma_\mu \, s_R\right) (x) \;\;,\;\;
\nonumber \\
{\cal O}_S(x)&\equiv& \left(\bar{d}_R \, s_L\right)(x)
 \left(\bar{d}_L \, s_R \right)(x) \, . 
\label{op}
\eea
Colour indices are summed inside brackets.
The factors $\eta_i(\mu)$ include the short-distance  QCD 
corrections from integrating out degrees of freedom between
the $W_R$ scale until some scale $\mu$ lower than the charm quark 
mass. They were calculated for the LR diagrams 
in \cite{eg} to one-loop and 
in \cite{giwi,ll} at two-loops for the LL part, 
we however only keep the one loop
value for consistency with the LR part and are functions
of the heavy-quark masses, gauge boson masses and $\alpha_S(\mu)$. 

Matrix elements of the Lagrangian in (\ref{eff}) are of course
scale independent. The scale $\mu$ dependence of the
$\eta_i(\mu)$ factors
cancels the one of the matrix elements of the operators
which are multiplying.
The dependence of  the hadronic matrix elements
on the renormalization scale $\mu$
just appears at next-to-leading order in 
$1/N_c$ which can at present be only estimated using models.
Below, we shall see that the leading order in $1/N_c$
estimate of the numerically dominant LR hadronic matrix 
element, which is model independent, 
has already a sizeable uncertainty.
We expect the matching scale $\mu$ to be a typical
hadronic scale around the rho meson mass.
In the case of the operators
${\cal O}_S(x)$ and ${\cal O}_{LR}(x)+
(2/N_c)\, {\cal O}_S(x)$, the present usage and uncertainties
of the leading in  $1/N_c$ estimate (see below),
make that varying the renormalization scale $\mu$ between
in $\eta_{4,5}(\mu)$ between  0.7 GeV and 1.2 GeV 
to be good enough for our purposes. 

Using $\overline m_c=(1.23\pm0.05)$ GeV \cite{mc}
for the $\overline{MS}$ pole charm-quark mass, 
$\Lambda^{(3)}_{QCD}=(350\pm100)$ MeV \cite{pdb} and 
the three-flavors one-loop running of $\alpha_S(\mu)$,
we get from \cite{eg} that $\eta_4(\mu)$
varies between 3.5 and 6.5 
and  $\eta_5(\mu)$ between 0.3 and 0.4.
These values depend on the right-handed
scale very weakly: varying $M_{W_R}$ from 1 to 10 TeV, 
they change  a couple of percent only. 
For the LL short-distance factors we take
\cite{ll}, 
$\eta_1(\mu)\simeq (1.0 \pm0.2) \alpha_S(\mu)^{-2/9}$, 
$\eta_2(\mu)\simeq (0.6 \pm0.1) \alpha_S(\mu)^{-2/9}$ and 
$\eta_3(\mu)\simeq (0.4 \pm0.1) \alpha_S(\mu)^{-2/9}$.

Let us now study the hadronic matrix elements needed.
For the LL operator we use the standard parametrization in terms
of the $\hat B_K$ parameter.
\bea
&&\langle K^0| {\cal O}_{LL}(x) | \bar{K}^0\rangle 
= \frac{4}{3} \alpha_S(\mu)^{2/9} \, \hat B_K\, f_K^2 m_K^2
\, .
\eea
The kaon decay coupling constant $f_K$ is 113 MeV \cite{pdb}
in this normalization. The hadronic matrix element $\langle K^0 | 
{\cal O}_{LL}(x) | \overline K^0 \rangle$
has been subject of much more work and
its present knowledge is summarized in \cite{burasorsay}.
We use the present favored range of values  $\hat B_K =
0.70 \pm 0.10$ \cite{burasorsay}.

The vacuum insertion approximation (VIA) has been used
generally in the literature to estimate LR hadronic
matrix elements. The same procedure gives $B_K=1$
at any scale. Unfortunately, the VIA is not a systematic expansion
in any parameter so that an estimate of the reliability 
of its predictions  and/or improving them  is not possible. 
We use instead the large $N_c$ expansion \cite{thooft} 
together with Chiral
Perturbation Theory (CHPT) (see \cite{toni} for a recent
 introductory
review and references) counting as organizative schemes. 
The combination of both techniques allows for a more systematic 
expansion and estimate of the uncertainties as we see below.
In the case of the LL operator, the leading
$1/N_c$ model independent result gives $\hat B_K=3/4$ 
with 0.25 as the  estimated uncertainty.
Using  the same expansion 
for the LR hadronic matrix elements, we get
\bea
&&\langle K^0| {\cal O}_S(x) | \bar{K}^0\rangle 
= \frac{\langle \bar{s}s + \bar d d\rangle^2}
{4f_K^2} + {\cal O}\left(\frac{m_K^2f_K^2}{N_c}\right) \, .
\label{me1} \\
&&\langle K^0| {\cal O}_{LR}(x) +\frac{2}{N_c} 
{\cal O}_S(x)|\bar{K}^0\rangle =  -f_K^2 m_K^2 
+ {\cal O} \left(\frac{\alpha_S}{\pi}\frac{\langle
\bar q q \rangle^2}{f_K^2} \right)\,,
\label{me2} 
\eea
where the quark condensates $\langle 0| \bar{s}s + 
\bar d d | 0 \rangle$  can be obtained from 
\bea
\langle 0 |
  \bar{s}s+\bar{d}d | 0 \rangle &=& -2 f_K^2\frac{m_K^2}{m_s+m_d}
(1-\delta_{K})\, .
\label{ss}
\eea
The parameter $\delta_{K}=0.35\pm0.10$ 
has been calculated using Finite Energy 
QCD sum rules \cite{narison}.
Using the $\overline{MS}$ running masses at 1 GeV, 
$m_s+m_d=(185\pm 30)$ MeV \cite{masses}, 
we obtain  
\be
\langle K^0| {\cal O}_S(x) | \bar{K}^0\rangle (\mu)
=(0.013\pm0.006) \hspace*{0.5cm} {\rm GeV}^4
\ee
and 
\be
\langle K^0| {\cal O}_{LR}(x)+\frac{2{\cal O}_S(x)}{N_c}
|\bar{K}^0\rangle (\mu) =-(3\pm3)\cdot 10^{-3}
\hspace*{0.5cm} {\rm GeV}^4,
\ee
where the scale $\mu$ varies between 0.7 GeV  and 1.2 GeV.
Due to the chiral structure of these operators, we
observe that the $1/N_c$ corrections to the matrix element
$\langle K^0| {\cal O}_S(x)|\overline{K^0}\rangle$ are
suppressed by a 
$m_K^2 f_K^4/\langle \bar q q\rangle^2$ factor. This makes
the main uncertainty in this matrix element operator to be the
one in the determination of the quark condensates, 
i.e. $\delta_K$. This translates into a 50 $\%$ uncertainty
in this matrix element, which can 
only be reduced with a more accurate
determination of the quark condensates.
The matrix element of the operator ${\cal O}_{LR}(x) + 
2 {\cal O}_S(x)/N_c$
has even larger relative uncertainties
due again to its chiral structure. In this case, 
there are non-factorizable 
$1/N_c$ corrections  which are not chirally suppressed, 
instead there is an additional $\alpha_S/\pi$ suppressing
factor. Fortunately, its leading
in $1/N_c$ contribution is chirally suppressed. In addition,
as we saw before,  the short-distance coefficient 
$\eta_5(\mu)$ is very small.
The discussion above makes clear that a large $N_c$ estimate of 
the matrix elements of the operators in the Lagrangian
in (\ref{eff}) is enough at present for this case.

Let us study the imaginary part of the $\Delta S=2$
efffective Lagrangian. For the 
left-handed sector we need the concurrence of the three-families
so that the dominance of the charm-charm box-diagrams is not true
anymore, in fact  the top-top box-diagram contribution 
dominates the CP-violating part of 
the $\Delta S=2$ effective Lagrangian. The dominance of
the charm-charm contributions is still true
for the imaginary part in the right-handed sector 
unless $\delta_1=\delta_2+n\pi$. Therefore, we 
can take the same effective Lagrangian in (\ref{eff}) unless 
$\delta_1=\delta_2+n\pi$. In that case, 
the right-handed sector behaves effectively as the left-handed
one in this respect, and no interesting bounds can be obtained.

In the approximations \cite{eduardo} where
\bea
\label{appro}
\Delta m_K\simeq2\mbox{\rm Re} M_{12}\;, \;\;\; \;\;\;\; \;\;\;
\mbox{with} \;\;\; \;\;\;\; \;\;\;
M_{12}\simeq-\langle K^0| {\cal L}_{eff}^{\Delta S=2} | 
\bar{K}^0\rangle /2m_K
 \eea
and
\bea
\epsilon_K\simeq\frac{1}{\sqrt{2}} 
e^{i \pi/4}\frac{\mbox{{\rm Im}}M_{12}}
{\Delta m_K^{\rm exp}}\, , 
\label{epsi}
\eea
where we have included in $\epsilon_K$ the long-distance
contributions to ${\rm Re} \, M_{12}$ in the experimental value of 
$\Delta m_K$, and 
using the effective Lagrangian in (\ref{eff}), we obtain 
\bea
\Delta m_K
\simeq \left[(0.40\pm0.20)-(4.5\pm2.5)\frac{g_R^2}{g_L^2}
\lambda_R \, 
\left(1-\frac{\dis \lambda_R^2}{\dis 2} \right)
 \, \cos(\delta_2-\delta_1) \, 
\left(\frac{1 \mbox{TeV}}{M_{W_R}}\right)^2
\right] \Delta m_K^{exp} \nonumber \\
\label{dm}
\eea
and
\bea
\label{elr}
\epsilon_K &\simeq& 
e^{i \pi/4} \left[ (2.7\pm 1.0) \cdot 10^{-3} \sin(\delta) 
\right. \nonumber \\ &-& \left.  
(1.6\pm0.9) \frac{g_R^2}{g_L^2} \lambda_R \, 
\left(1-\frac{\dis \lambda_R^2}{\dis 2}\right)\, 
\sin(\delta_2-\delta_1)
\left(\frac{1 \mbox{TeV}}{M_{W_R}}\right)^2 \right] \, ,
\eea
where the first contribution inside the squared brackets
is the LL contribution in each case, 
$\Delta m_K^{exp}=(3.491 \pm 0.014)\cdot 10^{-15}$ GeV 
 \cite{pdb}, and $|\epsilon_K^{\rm exp}| =(2.26 \pm 0.02) \cdot 10^{-3}$ 
\cite{pdb}. For the top quark-mass we have used the $\overline{MS}$
pole top-quark mass value $\overline m_t=(167\pm6)$ 
GeV \cite{topmass}. Notice that these two observables
only constrain two, $\delta$ and $\delta_1-\delta_2$,
out of the seven CP-violating phases we can have in the most 
general left-right model. The left-handed long-distance
contributions to $\epsilon_K$ (in our parametrization
of CP-phases) are expected to be negligible
since in this case the physics is  dominated by the
large top-quark mass contributions. 
Therefore, for the left-handed part,
the box-diagram is a good approximation in this case. There
are though, in principle, right-handed long-distance contributions 
to $\epsilon_K$ which are expected to be, as said before, of 
the same order as the right-handed short-distance ones. 
 
Let us now apply our results to left-right models
with different symmetries in the Lagrangian.
First we consider a manifestly LR symmetric 
model invariant under the transformation
(\ref{trans}) and with spontaneous breakdown of CP.
Remember that $g_R=g_L$ in these models.   
Diagonalization of the quark mass matrices for this case has been studied
in \cite{eg,frere}. 
In this type of models, $\lambda_R=\lambda$ and
all the relevant phases in the quark sector, namely 
$\delta$, $\delta_1$ and $\delta_2$ in our parametrization, 
are proportional to a single quantity 
$r=|\kappa_1/\kappa_2|\sin \varphi$ \cite{meie,frere}. Moreover,
when solving  for the quark masses one finds
the requirement $r<\mid m_b/m_t \mid$ 
\cite{meie,frere} which implies
the suppression of the phases $\delta$, $\delta_1$, and $\delta_2$ 
by this factor.
This particular feature makes this type of  models very 
predictive. The 
expressions of the phases $\delta$, $\delta_1$, and 
$\delta_2$ in terms of $r$, CKM matrix elements 
and quark masses can be found in \cite{meie}. 
The numerical analysis in \cite{meie} shows that the phases
$\delta$, $\delta_1$ and $\delta_2$ are actually very small
and $\cos(\delta_2-\delta_1)  \simeq 1$, so that 
there is no suppression in $\Delta m_K$ due to CP-violating
 phases. In this case and when the long-distance contributions 
are smaller than $\Delta m_K^{\rm exp}$, we have that
$(\Delta m_K)_{\rm box}$ has to be positive. Using this
positivity  one gets from (\ref{dm}) 
the following lower bound
\bea
M_{W_R} \gsim \left(1.6^{+1.2}_{-0.7}\right)\,
 \mbox{{\rm TeV}}\,.
\label{m1}
\eea 
This is quite general since only requires the natural 
expectation that long-distance contributions 
to $\Delta m_K$ are smaller than the experimental 
value. The assumption on the long-distance contributions 
done in \cite{beall} is similar to ours but the input values
and the hadronic matrix elements used are quite 
different. The fact
that the bound they get coincides with the central value 
in (\ref{m1}) is a numerical accident. (Notice 
that the left-handed part alone in 
\cite{beall} gives a 90 $\%$ contribution to 
$\Delta m_K^{\rm exp}$.)
The bound obtained in \cite{frere} is not valid
for models with $\cos (\delta_1 -\delta_2) \geq 0$
as for instance manifestly left-right models
we are considering here.
Another important difference of our analysis
respect to the ones in \cite{beall,frere}
is  the error bars.  They reflect the uncertainties
in the hadronic  matrix elements, and some input parameters,
mainly $\Lambda_{\rm QCD}$ as discussed above.
Taking into account realistic uncertainties, we  still
get a lower bound ($M_{W_R} \gsim 0.9$ TeV) from $\Delta m_K$
(together with  the assumption on the long-distance contributions 
explained above) which is larger than 
 the muon decay constraint we got
in Section \ref{third}, $M_{W_R}\gsim 549$ GeV,   
and the Tevatron direct limits $M_{W_R}\gsim 652$ GeV
\cite{tev}, $M_{W_R}\gsim 720$ GeV \cite{tev2}.
This shows the powerfulness of this low-energy
observable.

For manifestly LR models with no spontaneous CP
breaking, where CP-violation is parametrized 
by the complex Yukawa couplings,  
the transformation (\ref{trans}) requires $K_L=K_R$.
Therefore, only one independent CP-violating phase remains, 
the one analogous to the SM CKM phase, $\delta$.
The lower bound (\ref{m1})  holds in these models
since we have there $\delta_1=\delta_2=0$.

Finally, we analyse more general left-right
 models where we do not impose any discrete symmetry in the Lagrangian. 
As said before, in this case there are seven independent 
CP-violating  phases (one in the left-handed sector
and six in the right-handed in our parametrization).
They can have  both complex Yukawa (hard) origin and/or
spontaneous symmetry breaking origin and, in general,  
$\lambda_R \neq \lambda$. In the case where $\cos(
\delta_1-\delta_2) \geq 0$ and again long-distance
contributions are smaller than the measured $\Delta
m_K^2$, there is a lower bound on the $W_R$  mass
analogous to (\ref{m1}), which takes the form
\bea
M_{W_R}\gsim \left(3.4^{+2.5}_{-1.5}\right) \frac{g_R}{g_L}
\sqrt{ \lambda_R \, 
\left(1-\frac{\dis \lambda_R^2}{\dis 2} \right) \, 
\cos(\delta_2-\delta_1)}\;\, \mbox{{\rm TeV}}\,.
\label{m2}
\eea
If $\cos(\delta_1-\delta_2) \leq 0$ there is no cancellation
in (\ref{dm}) and to get lower bounds we need a stronger
assumption on the long-distance contributions, namely that they 
are smaller than the experimental value of $\epsilon_K$
and add positively, so in this case we have the constraint
$(\Delta m_K)_{\rm box} 
\leq \Delta m_K^{\rm exp}$. This case is actually
the one treated in \cite{frere}. Using this inequality
and (\ref{dm}) we get 
\bea
M_{W_R}\gsim \left(2.7^{+2.1}_{-1.2}\right) \frac{g_R}{g_L}
\sqrt{- \lambda_R \, 
\left(1-\frac{\dis \lambda_R^2}{\dis 2} \right) \, 
\cos(\delta_2-\delta_1)}\;\, \mbox{{\rm TeV}}\,.
\label{m2b}
\eea
Putting the worse case,  $\cos(\delta_2-\delta_1)=-1$,
and $\lambda_R=\lambda$ (as done in \cite{frere}), we get
$M_{W_R} \gsim (1.3^{+1.0}_{-0.6})$ TeV.
This lower bound updates the one in \cite{frere}.
Notice again that a realistic estimate of the present 
uncertainties allows, in this case, lower bounds
$M_{W_R} \gsim 0.7 $ TeV, just slightly
larger than direct Tevatron searches.

Let us turn to the analysis of the CP-violating parameter $\epsilon_K$.
>From (\ref{elr}) one notices that unless $(g_R/g_L)^2 \, 
\lambda_R \sin(\delta_1-\delta_2)$ is
close to zero, the experimental value for $\epsilon_K$
is saturated almost completely by the LL contribution, i.e. 
the LR contribution to $\epsilon_K$ 
for light $W_R$ gauge bosons is naturally larger than the LL one 
in general left-right models.
The reason was already given in the Introduction, it is
the need to have the three-families involved with 
non-degeneracy of the up-quark masses in the left-handed
sector,   while only the
two lightest are needed, in general,  in the right sector.
 This combined with the observed hierarchy of the left-handed CKM
angles\footnote{Left-handed CKM angles are measured in tree-level
processes, so we expect non-SM physics effects there to be negligible.}
 suppresses, in general,  the LL contribution 
to CP-violating parameters with respect to the LR one 
(in our parametrization).

Again, assuming long-distance contributions to
$\epsilon_K$ to be smaller than 
$\epsilon_K^{\rm exp}$, as noticed in \cite{meie}, 
allows to get relevant 
upper bounds on $M_{W_R}$ for most of the 
parameters space in general left-right models. 
For instance, if $ \sin (\delta) \leq 0.6$ in 
(\ref{elr}), we need a contribution larger or  
of the order of $\pm 10^{-5}  $ from the
right-handed part. We can from this observation  get a natural 
upper bound  on the $W_R$ mass in  general left-right
 symmetric models. So that we obtain the 
following upper bound
\bea
M_{W_R}\lsim {\cal O} \left(
350 \frac{g_R}{g_L}\sqrt{\lambda_R 
\, \left(1-\frac{\dis \lambda_R^2}{\dis 2} \right)
\, \mid \sin(\delta_2-\delta_1)\mid }
\right) \,\;  \mbox{{\rm TeV}}\,.
\label{m3}
\eea 
This bound will be violated unless 
$\sin(\delta_1-\delta_2)$ is close to zero.
 In that case,  the observed $\epsilon_K$ value 
has to be saturated by the left-handed contribution.
How close to zero depends on the lower bound values
for $M_{W_R}$.
Combining the lower bound in (\ref{m2}) and the upper
bound in (\ref{m3}),  we get that they are self-consistent unless
\be
\mid \sin ( \delta_1 - \delta_2 ) \mid \leq 10^{-4} \, ; 
\label{para}
\ee
i.e. $\mid \delta_1-\delta_2 \mid$ very close to 0, 
$\pi$ or $2 \pi$.
The upper bound in (\ref{m3}) 
 can be reduced by reducing the upper bound to
$\sin(\delta)$, of course. 

Let us now see how this bound applies to
the case of manifestly left-right symmetric models
with spontaneous CP violation. 
As was shown in \cite{meie}, in these type of models
$\mid \sin (\theta)\mid $ with $\theta=\delta$, 
$\delta_1$, or $\delta_2$ is typically of the
order of $10^{-2}$. Therefore 
the value of the phases in this type of
models fulfil the requirements for the bound
in (\ref{m3}) to hold in  most
of the parameter space. Using $\sin (\delta_1-\delta_2) 
\simeq$ $10^{-2}$ as a typical value of this region of
parameter space, we get
\bea
M_{W_R}\lsim {\cal O}(20)\, \mbox{{\rm TeV}}. 
\label{m22}
\eea 

A counter-example of the upper bound 
in (\ref{m3}) happens, for instance, in manifestly
symmetric left-right models with only complex Yukawa
couplings CP-violation or in models with $\delta_2=
\delta_1+n \pi$.

The bound in (\ref{m3}) is indicating that  
left-right symmetric models prefer (in general)
``light'' $W_R$ gauge bosons. Only very particular
 models, those where CP-violation requires also the
third family in the right-handed sector  can
naturally accommodate very heavy $W_R$ gauges boson masses.
This is a non-trivial bound which, for instance, constrains 
which type of CP-violation can have a model with a 
left-right symmetric scale of the order of $(10^7\sim
10^9)$ TeV, such as some left-right
symmetric models with a see-saw mechanism
for neutrino masses favored by neutrino physics \cite{neutrino}.

\section{Conclusions} 
\label{conclusions}
In this work we have studied and updated 
the bounds on the $W_R$ gauge boson mass and its mixing angle
$\zeta$ with its left-handed partner in general 
LR model.

Results independent on the quark sector assumptions 
and low energy QCD uncertainties  have been obtained 
by re-analysing LEP muon decay data. 
In the case the left-handed gauge coupling 
is equal to the right-handed one 
(for instance in manifestly left-right symmetric models), we get 
$M_{W_R}\geq 549$ GeV and $\mid \zeta\mid \leq 0.0333.$
Without that constraint, the best $\chi^2$ fit
to the muon decay data is obtained for $g_R/g_L=0.94\pm 0.09$
with $M_{W_R} \geq 485$ GeV and $\mid \zeta \mid \leq 0.0327$.

We have also considered bounds on the $W_R$ mass 
imposed by the neutral kaon observables, namely the
CP-conserving $\Delta m_K$ and the CP-violating
$\epsilon_K$. This has been done in left-right
models with different mechanisms of CP-violation. 
We have estimated the LR hadronic matrix elements and its
present uncertainty using a combined large 
$N_c$ expansion and CHPT analysis.  The uncertainty  
of the LR boxes contribution  to ${\cal L}^{\Delta S=2}_{\rm eff}$,
is due both to short-distance 
QCD corrections and hadronic matrix elements
contributions  and can be as large as $80\%.$ 
We want to emphasize that the low-energy
QCD dynamics is the main present uncertainty in
using the $ K^0 - \overline K^0$ system to constrain
left-right symmetric models parameters. 
We have included this uncertainty in our results.

With some explicitly given assumptions
on the  long-distance contributions
to $\Delta m_K$ and $\epsilon_K$ such like
that the long-distance contributions to a given observable
are smaller than its experimental value
(see Section \ref{fourth}), we get 
updated lower bounds on $M_{W_R}$ from the CP-conserving
$\Delta m_K$ for general left-right symmetric models
and strong upper bounds from the CP-violating $\epsilon_K$
parameter for most of the parameter space (\ref{para})
in general left-right models. The bounds obtained favor
quite light $W_R$ gauge bosons, namely masses below
a few hundreds  of TeVs 
(tens in the case of manifestly left-right models with spontaneous
CP violation).
This upper bound doesn't hold for the class of left-right
symmetric models where one needs the third family in the
right-handed sector. This can give a hint, in model building, 
of the CP-violating sector of the left-right symmetric models
where a very large left-right symmetric scale is required, 
as for instance models with a see-saw mechanism for the neutrino 
masses \cite{neutrino}. 
Upper bounds on the $W_R$ mass were obtained previously in 
\cite{frere} and  \cite{moha}. The hypothesis made 
in these references 
was however very strong, namely that  $\epsilon_K$ is
saturated completely by the right-handed contribution.
The  parameter space scenario where our upper
bound holds is much broader, only a small deviation
from the saturation of $\epsilon_K$ by 
the left-handed contribution  is enough. Our study
also clarifies which type of CP violation do
these upper bounds corresponds to.
Namely, 
the experimental value on
the CP-violating parameter $\epsilon_K$ likes to
have a quite light  intermediate left-right scale when
the right-handed third family is not required to contribute.

In particular, we have obtained the lower bound
$M_{W_R}\gsim(1.6^{+1.2}_{-0.7})$ TeV, 
in manifestly left-right symmetric models with either 
spontaneous and/or hard CP-violation,  
This bound complements
the one we got from the muon decay data in Section \ref{third}
and the direct Tevatron bounds \cite{tev,tev2}. 
For more general LR models (see Section \ref{fourth}
for details),  the lower bound we get is in (\ref{m2}).

This work shows the large potential and complementarity
of low-energy physics
to the direct searches in high-energy experiments.
In particular, in constraining
 new physics and/or  give hints on model building.
CP-violating quantities are very interesting in that respect 
due to its large suppression in the SM.
We have seen that only two CP-violating phases, out of
the seven in the most general left-right model, are
constrained from $\epsilon_K$, probably other 
low-energy CP-violating
observables give complementary information. 
One should also keep 
in mind that large CP-violating phases are welcome 
for baryogenesis \cite{baryo}. More work is needed, however,
in the low-energy QCD mastering in order to improve
the constraints we can get. 
The present hadronic uncertainties
dominate by far the error bars in the constrains we have obtained.
In particular, the inclusion of $(\Delta S=1)^2$
long-distance contributions, especially to $\Delta m_K$, 
will refine the constraints obtained here.

\subsection*{Acknowledgements}

We thank Jukka Maalampi, Toni Pich 
 and Arcadi Santamar\'{\i}a 
for clarifying discussions.
G.B. acknowledges the Spanish Ministry of
Foreign Affairs for a MUTIS fellowship and M.R. thanks the
Spanish Ministry of Education and Culture for a postdoctoral
fellowship at the Universitat de Val\`encia. 
This work has been supported in part by CICYT (Spain) under 
Grant No. AEN-96/1718.


\end{document}